\documentclass[aps,pre,superscriptaddress,twocolumn,amsmath,amssymb,showpacs]{revtex4}

\usepackage{amsmath}
\usepackage{graphicx}
\usepackage{amssymb}
\usepackage{dcolumn}
\usepackage{bm}

\makeatletter

\begin{document}

\title{Steering the potential barriers:  entropic to energetic}

\author{P.S. Burada} \thanks{Corresponding author: {\tt burada@pks.mpg.de}}
\affiliation{Max-Planck Institut f\"ur Physik komplexer Systeme,
  N\"othnitzer Str. 38, D-01187 Dresden, Germany}

\author{G. Schmid} \thanks{\tt gerhard.schmid@physik.uni-augsburg.de}
\affiliation{Institut f\"ur Physik,
  Universit\"at Augsburg, Universit\"atsstr. 1,
  D-86135 Augsburg, Germany}

\date{\today}


\begin{abstract}

We propose a new mechanism to alter the nature of the 
potential barriers when a biased Brownian particle under goes a 
constrained motion in narrow, periodic channel.
By changing the angle of the external bias, the nature of the
potential barriers changes from purely entropic to energetic which in turn
effects the diffusion process in the system.
At an optimum angle of the bias, the nonlinear mobility 
exhibits a striking bell-shaped behavior. 
Moreover, the enhancement of the scaled effective 
diffusion coefficient can be efficiently controlled by the angle of the bias. 
This mechanism enables the proper design 
of channel structures for transport of molecules and small particles. 
The approximative analytical predictions have been verified by 
precise Brownian dynamic simulations.

\end{abstract}

\pacs{05.60.Cd, 05.40.Jc, 02.50.Ey}

\maketitle

\section{Introduction}
\label{sec:intro}

The study of transport of molecules or small particles in 
narrow, confined systems such as pores, channels, or quasi-one-dimensional 
systems is of great interest due to it's major role in the process
of catalysis, particle separation and dynamical characterization of
these systems \cite{Hanggi_RMP,Burada_CPC,BM,PT}. 
Recent techniques enable to reveal the structural properties and
sequence of molecules like DNA or RNA when they pass
through narrow openings or the so-called bottlenecks, 
which cause entropic barriers \cite{Han,Keyser}.
These barriers are unavoidable and present in many fundamental 
processes that occur in nature, from molecules passing through 
ion channel proteins to catalytic process of ions in zeolites
or molecular sieves.
The nature of the potential barriers entirely depends
on the configurational space available for the particle,
i.e., whether the particle moving in an one-dimensional energetic
landscape or multi-dimensional confined geometry. 
In the former case, the nonlinear mobility increases
with the noise strength \cite{HTB, Reimann_PRL, Reimann_PRE,Lindner_FNL} 
due to noise assisted hopping over the potential barrier, and an 
opposite behavior can be observed in the later case \cite{Reguera_PRL}.
Consequently, the transport behavior of the 
particles/molecules sensitively depends on the nature of the 
potential barrier of the system.
Tuning the potential barriers is the key for 
controlling the transport of particles/molecules 
and allows for the development of new separation devices.

Here we analyze the transport and diffusion of point sized particles,
moving in two-dimensional narrow, confined geometries, and 
which are subjected to a constant bias. 
By changing the orientation of the bias the nature of the potential 
barriers of the system becomes controllable. 
The dynamics of the full system can be approximatively 
described by means of an one-dimensional 
kinetic equation, the so-called Fick-Jacobs equation,  
which contains an effective potential function
which subsumes/incorporate the geometrical restrictions 
\cite{Zwanzig, Reguera_PRE, siwy,Ilona, berzhkovski,Reguera_PRL}. 

The paper is organized as follows. In the next section we 
introduce the model system, namely the dynamics of a Brownian
particle in a confined geometry with irregular boundaries. 
A simplified approximative one-dimensional kinetic description 
for the full system is discussed in section~\ref{sec:1Dmodeling}.
An analytical treatment of the problem for finding the main transport
characteristics is presented in section~\ref{sec:trans}.
The numerical techniques and main findings are discussed in section~\ref{sec:results}.
We present our main conclusions in section~\ref{sec:conclusions}.

\section{Confined Brownian motion}
\label{sec:model}

The overdamped dynamics of a Brownian particle, in a confined geometry, 
subjected to the constant force $\vec{F} $ can be described by means of the Langevin
equation written as

\begin{subequations}
  \label{eq:langevin}
\begin{align}
  \gamma \, \frac{\mathrm{d}\vec{r}}{\mathrm{d} t} = &
  \,\vec{F}
  +\sqrt{\gamma \, k_{\mathrm{B}}T}\, \vec{\xi}(t)\, ,
\intertext{with}
\vec{F} = & F_\parallel \vec{e}_{x} - F_\perp \vec{e}_{y}
\end{align}
\end{subequations}
where $F_\parallel = |\vec{F}|  \cos(\theta)$ and $F_\perp = |\vec{F}| \sin(\theta)$
denote the force components along and  perpendicular to the 
2D channel direction, respectively.
$\vec{r}$ denotes the position of the particle in 2D, $\gamma$ is
the friction coefficient and
$\vec{\xi}(t)$ is the Gaussian white noise with zero mean and correlation function: 
$\langle \xi_{i}(t)\,\xi_{j}(t') \rangle = 2\, \delta_{ij}\, \delta(t - t')$
for $i,j = x,y$. 

\begin{figure}[t]
  \centering
  \includegraphics{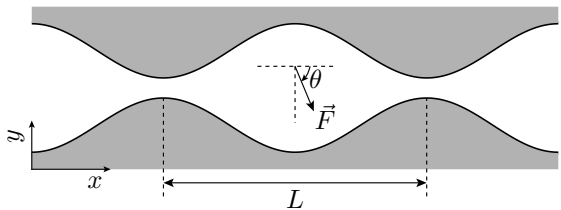}
  \caption{Schematic illustration of the periodic channel confining the
    movement of a Brownian particle which is subjected to a constant
    force $\vec{F}$. $L$ is the periodicity of the channel structure
    and $\theta$ is the angle between the force vector and the
    positive $x$-axis.}
  \label{fig:tube}
\end{figure}

In the presence of confinement, this equation has to be solved by
imposing reflecting (no-flow) boundary conditions
at the walls of the structure. For the 2D structure depicted in
Fig.~\ref{fig:tube}, the shape of the symmetric boundary is defined
by the function
\begin{align}
  \label{eq:shape}
  \omega(x) = a\sin(2\pi x/L) + b\,,
\end{align}
where $L$ is the periodicity of the channel and
the parameters $a$ and $b$ control the slope and the channel width at 
the bottleneck. 
Here, the condition $b>a$ should be satisfied in order to enable
the passage of  particles from one cell to the other.
The sum and difference of the two parameters $a+b$ and
$b-a$ yield  half of the maximum width and half of the minimum width of the
channel, respectively.
The local width of the channel is given by $2\, \omega(x)$.
For the sake of simplicity, we use dimensionless units. In
particular, we scale lengths by the periodicity of the channel $L$,
time by $\tau=\gamma L^2/k_{\rm B}T_{\rm R}$, the corresponding characteristic
diffusion time at an arbitrary but irrelevant reference temperature $T_{\rm R}$, 
and force by $F_{\rm R} = \gamma L/\tau$.

In dimensionless form the Langevin-equation~\eqref{eq:langevin} and
the boundary functions \eqref{eq:shape} read
\begin{subequations}
\begin{align}
  \label{eq:dllangevin}
  & \frac{\mathrm{d}\vec{r}}{\mathrm{d} t} = F_\parallel \vec{e}_{x} - F_\perp  \vec{e}_{y}
  +\sqrt{D}\, \vec{\xi}(t)\, ,\\ 
  \label{eq:dlshape}
  & \omega(x) = a\sin(2\pi x) + b \,,
\end{align}  
\end{subequations}
with the rescaled temperature $D=T/T_{\rm R}$.

\section{Approximative 1D kinetics}
\label{sec:1Dmodeling}

The dynamics of the system can be described by means of
a kinetic equation, obtained originally from the 2D 
Smoluchowski equation after elimination of the $y$ coordinate.  
The resulting equation, namely the Fick-Jacobs (FJ) equation, 
with a spatially dependent diffusion coefficient, 
reads for $|{\omega}^\prime (x)| \ll 1$ 
(the prime refers to the derivative with respect to $x$)
\cite{Zwanzig}
\begin{align}
\label{eq:fickjacobs}
\frac{\partial}{\partial t} P(x,t) =
\frac{\partial}{\partial x} \,{\cal D}(x) \,e^{-A(x,\theta)/D}
\frac{\partial}{\partial x} e^{A(x,\theta)/D} P(x,t) \,.
\end{align}
Here, $P(x,t)$ is the probability distribution function 
along the length of the 2D channel,
${\cal D}(x)$ is the spatially diffusion coefficient \cite{Reguera_PRE} which reads
\begin{align}
{\cal D}(x)= \frac{D}{(1+\omega^\prime(x)^2)^\alpha} \, ,
\end{align}
where $\alpha = 1/3$.
Therefore, the boundary function given by Eq.~\eqref{eq:dlshape} determines ${\cal D}(x)$.
Introducing the spatially diffusion coefficient into the 
kinetic equation, improves the accuracy, and extends it's validity to 
more winding structures \cite{Reguera_PRL, Zwanzig, Reguera_PRE,Kalinay}.

In Eq.~\ref{eq:fickjacobs},  $A(x)$ is the free energy, 
which in presence of a constant bias reads,  cf. Ref. \cite{Burada_PRL, Burada_EPJB},
\begin{subequations}
\label{eq:effpotential}
\begin{align}
  & A(x,\theta)= -F_\parallel \,x - D\, \ln\left[2\,\omega(x)
    \frac{\sinh\epsilon}{\epsilon}\right] \\
  \intertext{with}
  &  \epsilon = \frac{F_\perp \,\omega(x)}{D} \,.
\end{align}
\end{subequations}

Note that the free energy $A(x,\theta)$
can be tuned by altering the aspect ratio $\epsilon$,
which is a function of the key parameters, namely, 
the angle of the applied bias, 
the noise strength $D$ and the shape of the periodic potential.
Thus, it is instructive to analyze the two limiting situations for 
the free energy $A(x,\theta)$ that can be obtained depending on 
the angle of the bias.

\subsection{Bias in the channel direction}

For $\theta = 0$, i.e., in the case of a constant force pointing in the channel direction, 
the dimensionless free energy function, 
Eq.~\ref{eq:effpotential}, modifies into the form
\begin{align}
  \label{eq:Ent}
  A(x) = - |\vec{F}| x -D\,\ln \left[2\,\omega(x)\right]
\end{align}
resembling a potential function with purely entropic barriers
\cite{Burada_CPC, Reguera_PRL, Burada_PRE,Burada_BioSy,ai2006, Burada_EPL}.

\subsection{Bias perpendicular to the channel direction}

For $\theta = \pi/2$, Eq.~\ref{eq:effpotential} turns into
\begin{align}
  \label{eq:thpib2}
  A(x) = - D\, \ln\left[\frac{2 D}{|\vec{F}|}\, \sinh{\left(\frac{|\vec{F}| \omega(x)}{D}\right)}\right]\,.
\end{align}
Remarkably, this effective potential
depends on the absolute value of the bias, 
the noise strength and the geometry of the structure 
in a non-trivial way \cite{Burada_PRL}.
Interestingly, this effective potential exhibits energetic barriers persisting even in the limit of 
small noise, i.e., for $D \to 0$. 

Thus, by changing the angle of the applied bias one can effectively tune
the nature of the potential barriers, i.e., from entropic to energetic.
If the direction of the applied bias is parallel to the direction
of the channel, the free energy function will have a purely 
entropic nature, where the barrier height is a function of the noise strength.
At a finite angle of the applied bias, the free energy function will have both
an entropic and an energetic contribution. At small noise strengths it is 
purely energetic and at high noise strengths it is purely entropic \cite{Burada_EPJB}.

\section{Transport characteristics}
\label{sec:trans}

For calculating the key quantities of transport like the average particle current, 
or equivalently the nonlinear mobility, and the effective diffusion coefficient
we use the mean first passage time (MFPT) approach.
For any finite force the average particle current in periodic
structures can be obtained from 
\begin{align}
  \label{eq:parcur}
  \langle \dot{x} \rangle = \frac{1}{T_{1}(x_0 \to x_0+1)} \,,
\end{align}
where $T_{1}(x_0 \to x_0+1)$ is the first moment of the
first passage time distribution for reaching 
$x = x_0+1$ starting out at $x = x_0$.
Note that the mean first passage time diverges for a 
vanishing force and consequently leads to a vanishing current.
The effective diffusion coefficient is defined as the asymptotic
behavior of the variance of the position and is given by \cite{Reimann_PRL, Reimann_PRE, Lindner_FNL}
\begin{align}
  \label{eq:diffcoef}
  D_{\mathrm{eff}} = \frac{T_2(x_0 \to x_0+1) - \left[ T_1(x_0 \to x_0 +1) \right]^2}
  {2 \left[T_1(x_0 \to x_0+1)\right]^3} \,,
\end{align}
where $T_2$ is the second moment of the distribution of the first passage times which can be calculated 
by Eq.~\eqref{eq:st-average}.

The $n$th moment of the first passage time distribution 
can be determined recursively by means of
\begin{align}
  \label{eq:st-average}
  T_n (x_0 \to x_0+1)&  := \langle t_n(x_0 \to x_0+1)\rangle
  \nonumber \\
   & \hspace*{-1cm}= n \int_{x_0}^{x_0+1}   
  \frac{e^{A(x,\theta)/D}}{{\cal D}(x)}\,\mathrm{d}x
  \int_{-\infty}^{x}
  e^{-A(y,\theta)/D} \,\mathrm{d}y
   \nonumber \\
  & \hspace*{-0.7cm}\times T_{n-1}(y \to x_0+1) 
\end{align}
for $n \in \mathbb{N}$ with $T_0(x_0 \to x_0+1) = 1$.
Since  the spatially dependent diffusion coefficient is periodic,
i.e., ${\cal D}(x+1) = {\cal D}(x)$, and $A(x+1,\theta) = A(x,\theta) - F_\parallel$, 
the recurrence relation can be simplified.
The nonlinear mobility is obtained (by setting $x_0 = 0$), 
\begin{align}
  \label{eq:avg-curr}
  \mu = \frac{\langle \dot{x} \rangle}{F_\parallel} = \frac{1-e^{-F_\parallel/D}}
  {F_\parallel \,\displaystyle \int_{0}^{1} I(x,\theta)\,\mathrm{d}x} \,,
\end{align} 
and the effective diffusion coefficient is
\begin{align}
  \label{eq:eff-diff}
  D_{\mathrm{eff}} = \frac{\displaystyle 
    \int_{0}^{1} \int_{x-1}^{x}
    \frac{{\cal D}(z)}{{\cal D}(x)}\,
    \frac{e^{A(x,\theta)/D}}{e^{A(z,\theta)/D}}\,[I(z,\theta)]^2
    \,\mathrm{d}x \,\mathrm{d}z}
  {\left[\displaystyle \int_{0}^{1} 
      I(x,\theta)\,\mathrm{d}x \right]^3} \,,
\end{align}
where the integral function $I(x,\theta)$ reads
\begin{align}
  \label{eq:int-fun}
  I(x,\theta) = \frac{e^{A(x,\theta)/D}}{{\cal D}(x)}
  \displaystyle \int_{x-1}^{x} e^{-A(y,\theta)/D} \,\mathrm{d}y \,.
\end{align}

\subsection{Deterministic limit}

In the deterministic limit, i.e., $D \rightarrow 0$, 
there is no entropic contribution and the free energy function
modifies into $A(x,\theta)= -F_\parallel\,x - F_\perp \omega(x)$, which corresponds
to the washboard potential. The critical angle for which the potential barriers disappear is given by
\begin{equation}
\label{eq:critical}
{\theta}_c = \mathrm{arctan}\left(\frac{1}{2\pi a}\right) \,.
\end{equation} 

Then, the asymptotic value of the nonlinear mobility reads 
\cite{Risken, Burada_CP}
\begin{equation}
  \label{eq:zeroV}
  \lim_{D \to 0}\,\mu = \left\{ \begin{array}{lcr}
      \sqrt{1 -\left[ \frac{\tan(\theta)}{\tan({\theta}_c)} \right]^2} & \theta < {\theta}_c \\
      0 & \theta \geq {\theta}_c
    \end{array}\right.
\end{equation}
and the corresponding effective diffusion coefficient reads
\begin{equation}
  \label{eq:zeroDeff}
  \lim_{D \to 0}\,\frac{D_{\mathrm{eff}}}{D} = \left\{ \begin{array}{lcr}
      \displaystyle \frac{1}{\sqrt{1 -\left[ \frac{\tan(\theta)}{\tan({\theta}_c)} \right]^2}} 
      & \theta < {\theta}_c \\
      0 & \theta \geq {\theta}_c
    \end{array}\right. 
\end{equation}

\subsection{Strong noise limit}

In the strong noise limit, i.e., $D \rightarrow \infty$, 
the entropic contribution is more dominant in the free energy function,
$A(x)= -F_\parallel \,x - D \ln[2 \omega(x)]$.
Then, the asymptotic values of the nonlinear mobility and the scaled effective diffusion
coefficient read \cite{Lifson-Jackson,Burada_PRE}
\begin{subequations}
  \label{eq:highD}
\begin{align}
  \lim_{D \to \infty}\,\mu &= \frac{1}
  { \displaystyle \int_{0}^{1} \frac{D}{{\cal D}(x)\, \omega(x)} \,\mathrm{d}x \,
    \int_{0}^{1}\omega(y) \mathrm{d}y} \\[0.5ex]
  & = \lim_{D \to \infty} \frac{D_{\mathrm{eff}}}{D} 
\end{align}
\end{subequations}

\subsection{Zero bias limit: along the longitudinal direction}

In the absence of force component along the transversal direction, 
i.e., for the case of $\theta = \pi/2$, one can use the Lifson-Jackson
analytical formula for the scaled effective diffusion coefficient which 
is given by \cite{Lindner_FNL,Lifson-Jackson,Risken,Burada_PRE}

\begin{align}
\label{eq:Lif-Jac}
\lim_{\theta \to \pi/2} \frac{D_{\rm eff}}{D} = \frac{1}
	{\displaystyle \int_{0}^{1}\frac{D\,e^{A(x,\theta)/D}}{{\cal D}(x)}  \mathrm{d}x
	\int_{0}^{1}e^{-A(y,\theta)/D} \mathrm{d}y}\,.
\end{align}

As the longitudinal force component is zero for $\theta =\pi/2$, 
the nonlinear mobility vanishes.

\section{1D {\it vs.} 2D modeling }
\label{sec:results}

Analytically derived transport characteristics have been compared with
the results obtained from the Brownian dynamic simulations, 
performed by a numerical integration of the Langevin equation, 
Eq.~\ref{eq:langevin}, using the standard stochastic Euler-algorithm. 
The average particle current is defined as
\begin{align}
\label{eq:velocity}
\langle \dot{x} \rangle = \lim_{t\to \infty} \frac{x(t)}{t} \,, 
\end{align}
and the corresponding effective diffusion coefficient as, 
\begin{align}
\label{eq:diff}
D_{\mathrm{eff}} = \lim_{t \to \infty} \frac{\langle
x^{2}(t) \rangle - \langle x(t) \rangle^{2} }{2 t} \,.  
\end{align}
For the calculations the geometrical parameters are set to
$a = 1/2\pi$ and $b=1.02/2\pi$.

\begin{figure}[t]
  \includegraphics{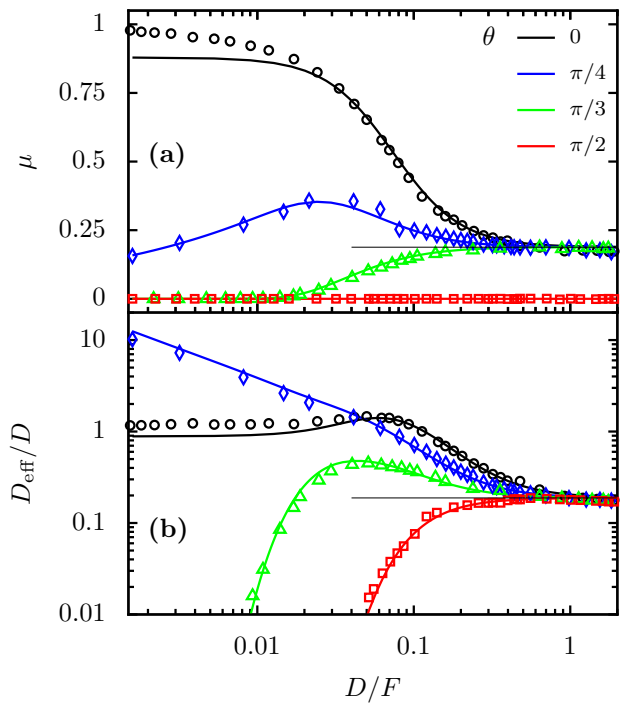}
  \centering
  \caption{(Color online). 
    The nonlinear mobility $(a)$ and the
    effective diffusion coefficient $(b)$ as function of the 
    scaling parameter $D/F$ ($F = |\vec{F}|$) 
    for various angles of the force.
    The lines correspond to the analytical predictions, 
    $(a)$ Eq.~\ref{eq:avg-curr} and $(b)$ Eq.~\ref{eq:eff-diff}, whereas
    the symbols correspond to the results of the Langevin dynamic simulations 
    for the two-dimensional structure with the shape defined by the 
    dimensionless function $\omega(x) = \left[\sin(2\pi x) + 1.02\right]/2\pi$.
    In limit $D \to \infty$, both $\mu$ and $D_{\mathrm{eff}}/D$ reach a 
    limiting value of 0.188 (thin black line) predicted by Eq.~\eqref{eq:highD}.}
  \label{fig:cur-diff}
\end{figure}

\begin{figure}[ht]
  \includegraphics{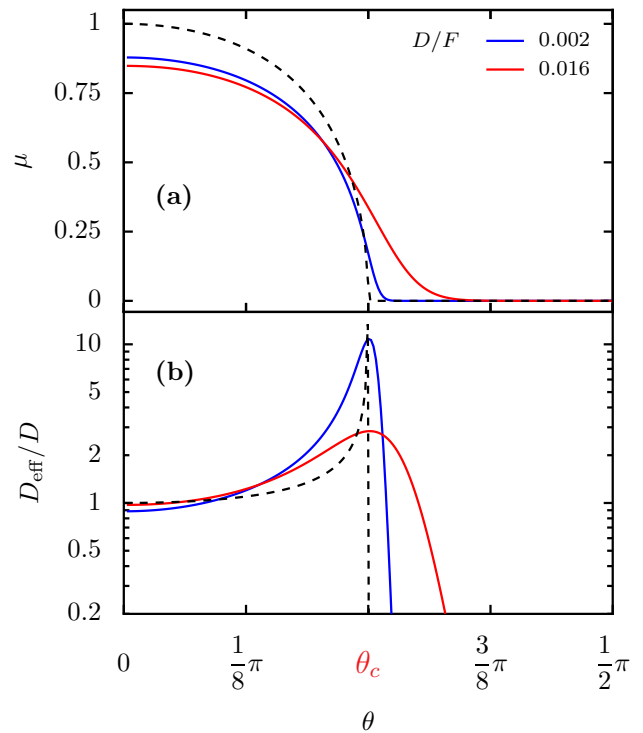}
  \centering
  \caption{(Color online). Analytically obtained nonlinear mobility, Eq.~\eqref{eq:avg-curr}, 
    and the scaled effective diffusion coefficient, Eq.~\eqref{eq:eff-diff},   
    as function of the rotation angle $\theta$.
    The asymptotic behaviors of $\mu$, Eq.~\eqref{eq:zeroV}, and 
    $D_{\mathrm{eff}}/D$, Eq.~\eqref{eq:zeroDeff}, in the limit $D \to 0$ are plotted in dashed lines. 
    In the deterministic limit, both $\mu$ and $D_{\mathrm{eff}}/D$ become zero for $\theta \geq {\theta}_c$.
    Here, we present only the analytical predictions for clarity, whereas the agreement with
    numerical simulations is same as in Fig.~\ref{fig:cur-diff}.}
\label{fig:theta}
\end{figure}

The nonlinear mobility $(a)$
and the scaled effective diffusion coefficient $(b)$ as function of the 
scaling parameter $D/F$ for different 
angles of the bias are depicted in Fig.~\ref{fig:cur-diff}.
For $\theta = 0$, i.e., for a purely entropic potential, 
the nonlinear mobility decreases 
monotonically with increasing the scaling parameter $D/F$, whereas 
the scaled effective diffusion coefficient $D_\mathrm{eff}$
exceeds the bulk diffusion $D$. 
Interestingly, for $\theta = \pi/4$ the nonlinear mobility
shows a bell-shaped behavior, exhibiting a peak over a range
of $D/F$ values. 
It is due to the fact, that for very small $D/F$ values 
the energetic nature of the barriers dominates resulting an increase in $\mu$.
Contrarily, for large values of the scaling parameter 
$D/F$ the entropic nature dominates, which
slows down $\mu$. 
For intermediate values of $D/F$ the potential function contains
both the entropic and energetic contribution resulting a peak in $\mu$.
Moreover, the scaled effective diffusion shows also a peculiar behavior for $\theta=\pi/4$.
For small $D/F$ values, the scaled effective diffusion is greatly enhanced and 
drops down to the bulk diffusion constant when the scaling parameter $D/F$ is increased.
Note that a similar monotonic increase of the scaled effective 
diffusion coefficient with decreasing noise level was 
observed in periodically segmented channel systems \cite {Marchesoni, BerezhkovskiiAM}.
For $\theta = \pi/2$, $F_\parallel = 0$, and as one would expect 
$\mu = 0$, and $D_\mathrm{eff}/D$ exhibits a monotonically increasing behavior with the scaling parameter, 
which could also be predicted by the Lifson-Jackson formula, Eq.~\eqref{eq:Lif-Jac}. 
Note that a similar behavior of the scaled effective diffusion coefficient was 
observed in periodically segmented channel systems \cite {Marchesoni, BerezhkovskiiAM}.

It is worth to mention that the agreement between the results of the 1D modeling  (lines)
and the numerical simulations of the full 2D-system (symbols)
is quite good, and the little deviation at small $D/F$ for $\theta=0$
is a known failure of the FJ approximation \cite{Burada_CPC, Burada_PRE},
where the assumption of equilibration along the transversal direction does not completely hold. 

In Fig.~\ref{fig:theta} the dependence of the nonlinear mobility and 
the scaled effective diffusion coefficient on the angle of the applied bias is depicted. 
With increasing angle, the nonlinear mobility monotonically decreases. 
It is the increase of the energetic barrier height of the potential function with $\theta$
which leads to the continuous deterioration of the nonlinear mobility depicted in Fig.~\ref{fig:theta}a. 
In addition, the diffusion coefficient shows a striking behavior: 
with increasing height of the energetic barriers, the scaled effective diffusivity initially goes up, 
reaches a maximum at the critical tilt, at $\theta = \theta_c$, then drops down. 
This is in accordance to the giant enhancement of diffusion 
phenomena found for tilted periodic energetic potentials \cite {Reimann_PRL, Reimann_PRE}. 
The maximum value depend strongly on the dimensionless scaling parameter $D/F$. 
With decreasing $D/F$ the peak of the scaled effective diffusion coefficient gets more pronounced, 
cf. the giant diffusion phenomena \cite{Reimann_PRL, Reimann_PRE}.
In the deterministic limit, i.e., for $D/F \to 0$, $D_{\mathrm{eff}}/D$  
shoots up at the critical angle of the rotation.
But above the critical angle (${\theta}_c$) both the nonlinear mobility and 
the scaled effective diffusion coefficient drops down to zero, see Eq.~\ref{eq:zeroV} and ~\ref{eq:zeroDeff}. 

In other words, by tuning the angle of the bias, 
the diffusion of particles inside the channel can be enhanced by 
many orders of magnitude.
This drastic enhancement in $D_{\rm eff}/D$ was also observed in segmented 
channels \cite{Borromeo}. 

\begin{figure}[t]
  \includegraphics{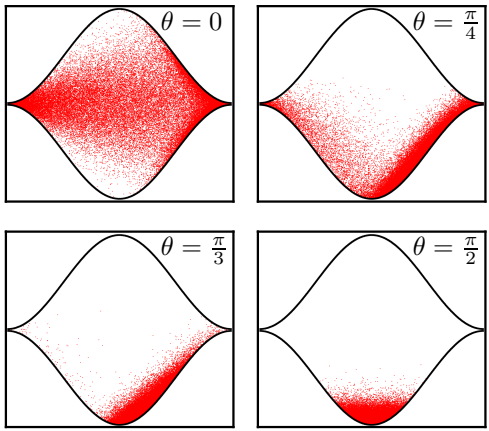}
  \caption{(Color online). The numerically obtained steady-state 
    distribution of particles in a two-dimensional channel with the shape defined by the 
    dimensionless function $\omega(x) = \left[\sin(2\pi x) + 1.02\right]/2\pi$ 
    for different angles of the bias and at the scaling parameter $D/F = 0.016$ ($F = |\vec{F}|$).}
    \label{fig:snap}
\end{figure}

\subsection{Steady-state distribution}

The steady-state distribution of the particles in a two-dimensional geometry is presented in
Fig.~\ref{fig:snap} for different angles of the applied bias.
For $\theta = 0$ only the longitudinal force component survives,
forcing the particles to move from one cell to the other. With increasing angle,
the strength of the longitudinal force component decreases and the transversal component 
increases, as a result particles are forced to move along the 
lower boundary of the channel. 
At higher angles ($\theta \to \pi/2$), the longitudinal force tends to zero, the particle 
flows in forward and backward direction are balanced,  and as a result the net motion becomes zero.

\begin{figure}[ht]
  \centering
  \includegraphics{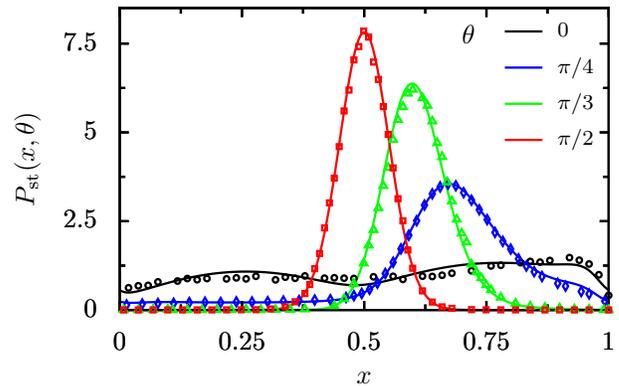}	
  \caption{(Color online). Normalized steady-state 
    probability density of particles along the propagation direction 
    for various angles of the bias ($F = |\vec{F}|$).
    Lines correspond to the analytical prediction, Eq.~\eqref{eq:steadystate},
    and symbols correspond to the results of the Langevin simulations of the full system.
    The other parameters are same as in Fig.~\ref{fig:snap}.}
  \label{fig:steadystate}
\end{figure}

For the reduced one-dimensional kinetic equation, Eq.~\eqref{eq:fickjacobs},
the normalized steady-state probability density 
$P^{\mathrm{st}}(x, \theta)$ along the propagation direction 
can be derived analytically, and is given by \cite{Burada_PRE}
\begin{align}
  \label{eq:steadystate}
  P^{\mathrm{st}}(x, \theta) = 
  \frac{e^{-A(x,\theta)/D} \displaystyle \int_{x}^{x+1} 
    \frac{e^{A(y,\theta)/D}}{{\cal D}(y)}\, \mathrm{d}y} 
  {\displaystyle \int_{0}^{1} \frac{e^{A(x,\theta)/D}}{{\cal D}(x)}
   \int_{x-1}^{x} e^{-A(y,\theta)/D} \,\mathrm{d}y }\,.
\end{align}
A comparison with the numerically obtained probability 
density, is depicted in Fig.~\ref{fig:steadystate}
for various angles of the bias. 
At $\theta = 0$ only the longitudinal force component survives,
forcing the particles move from one cell to the other, 
results in a flat distribution of $P_{\mathrm st}(x,\theta)$
over the full length of the cell.
As the angle increases, the transversal force component 
increases and the particles are more focused towards the middle of the cell.
For $\theta = \pi/2$, there is no force component acting along
the longitudinal direction, and particles settle down at the bottom of the cell, 
see Fig.~\ref{fig:snap}.
The finite half-width in the probability distribution
is due to the presence of noise in the system.
Surprisingly, the dynamics of the full system is almost captured 
within the approximative FJ description, whereas 
the small deviation at  $\theta = 0$ 
is a known failure of FJ approximation \cite{Burada_PRE, Kalinay}.

\section{Conclusions}
\label{sec:conclusions}

We have shown that the transport nature of the Brownian particles in a channel 
with periodically varied cross-section area can 
effectively be tuned by altering the angle of the applied bias.
When the direction of the applied bias is parallel to the direction
of the channel, the effective potential will have a purely entropic nature.
In this situation, the nonlinear mobility decreases monotonically with 
increasing scaling parameter $D/F$.
For $\theta \approx \pi/4$, $\mu$ exhibits a bell-shaped
behavior which is resulting from the competition of the 
entropic and energetic contributions to the free energy profile. 
For larger values of the angle, the energetic nature of the system rules the dynamics and
a monotonic increase of $\mu$ with $D$ is observed.
Interestingly, by rotating the bias from parallel to perpendicular to the channel 
direction, the scaled effective diffusion initially increases,
reaches the maximum at the critical angle of the bias, 
which depends on the amplitude of the geometric structure, and then drops down to zero. 
Summing up, by changing the angle of applied bias the diffusive transport of the particles
can effectively be controlled. We have shown that the dynamics of 
the full system can be described adequately within 
a reduced one-dimensional kinetic description.

This new approach of controlling the transport of Brownian particles in
narrow confined structures has wide range of applications, including
catalysis, particle separation, 
fluid mixing \cite{Frommelt}, and many more.

\subsection*{ACKNOWLEDGMENTS}

This work has been supported by
the Max Planck society and
Volkswagen foundation project I/83902.

\end{document}